\documentclass[aps,prl,twocolumn,superscriptaddress,showpacs]{revtex4}
\usepackage{graphicx}
\usepackage{dcolumn}
\usepackage{bm}

\begin{document}

\title{Contact Dependence of Carrier Injection in Carbon Nanotubes:
       An {\em Ab Initio} Study}

\author{Norbert Nemec}
\affiliation{Institute for Theoretical Physics,
             University of Regensburg,
             D-93040 Regensburg, Germany}

\author{David Tom\'{a}nek}
\affiliation{Physics and Astronomy Department,
             Michigan State University,
             East Lansing, Michigan 48824-2320 }

\author{Gianaurelio Cuniberti}
\affiliation{Institute for Theoretical Physics,
             University of Regensburg,
             D-93040 Regensburg, Germany}

\date{23 February 2006}

\begin{abstract}
We combine {\em ab initio} density functional theory with
transport calculations to provide a microscopic basis for
distinguishing between good and poor metal contacts to
nanotubes. Comparing Ti and Pd as examples of different contact
metals, we trace back the observed superiority of Pd to the nature
of the metal-nanotube hybridization. Based on large scale
Landauer transport calculations, we suggest that the
`optimum' metal-nanotube contact combines a weak hybridization
with a large contact length between the metal and the nanotube.
\end{abstract}




\maketitle



A major challenge linked to the use of carbon
nanotubes~\cite{iijima-hmogc1991} in future electronic devices is
to understand the profound effect of the nanotube-metal contact on
transport. Weak nanotube-metal coupling, found in nanotubes
deposited on metal electrodes, has been shown to cause Coulomb
blockade behavior~\cite{tans-iscnaqw1997}. In spite of
significant progress in maximizing the contact area by depositing
metal on top of nanotubes~\cite{bockrath-resbdiscn2001}, the
transparency of such contacts exhibits strong sample-to-sample
variations and depends strongly on the contact metal. Reports of
low contact resistance between nanotubes and Au or
Au/Cr~\cite{yaish-enoscnuaafm2004,liang-f-piianew2001} are in
stark contrast to the high resistance observed in nanotube
contacts with Au/Ti~\cite{wakaya-cromcn2003}. The transparency of
Pd-based contacts has been reported as superior in comparison to
using Ti, Pt and Al as contact
metals~\cite{javey-bcnft2003,mann-btimnwrpoc2003,chen-tromcitpocnft2005}.
Additional modulation of the Pd-nanotube contact transparency has
been reportedly achieved by modulating the gate
voltage~\cite{babi-oofriscn2004}. Reports suggesting that carrier
injection occurs only at the edge of the contact
region~\cite{mann-btimnwrpoc2003} appear to contradict the
observed dependence of the contact resistance on the length of the
contact~{\cite{wakaya-cromcn2003}}.

Published theoretical results include studies of the electronic
structure at a nanotube-Au interface and transport properties of a
nanotube-Al junction~\cite{nardelli+rubio}. {\em Ab initio}
calculations furthermore suggest that Ti contacts may be superior
to those with Al or Au~\cite{palacios-fptimnwrc2003}, and that the
Schottky barrier between semiconducting tubes and Pd is lower than
with Au or Pt {\cite{shan-aisosbamc2004}}. Due to the limitation
to specific contact geometries and small system dimensions,
however, general trends are hard to extract, and an extrapolation
to experimentally relevant system sizes is difficult.

Here we combine {\em ab initio} electronic structure studies with
large scale transport calculations to gain microscopic insight
into the relative importance of the interface morphology, the type
of the contact metal, and the length of the contact region when
optimizing the metal-nanotube contact. {\em Ab initio} density
functional studies were used to determine the charge
redistribution and electrostatic potential in the contact region.
In a second step, the electronic structure results were mapped
onto a model tight-binding Hamiltonian
suitable for transport
calculations. We found that transmission is maximized in the case
of weak metal-nanotube coupling, exhibited by extended Pd contacts.

\begin{figure}[b]
\includegraphics[width=0.90\columnwidth]{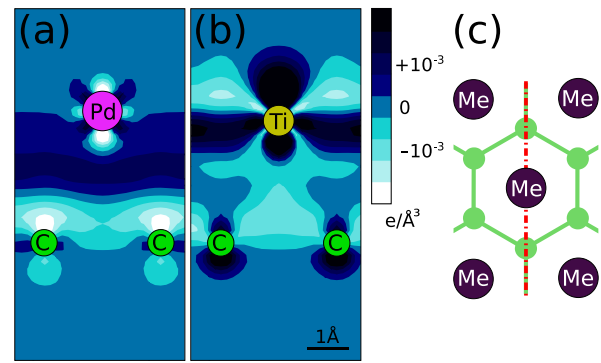}
\caption{(Color online) Charge density redistribution
$\Delta{\rho}(\textbf{r})= \rho_{\text{\textrm{Me/C}}}(\textbf{r})
-\rho_{\text{\textrm{Me}}}(\textbf{r})-
\rho_{\text{\textrm{C}}}(\textbf{r})$ in (a) Pd and (b) Ti
monolayers interacting with a graphene layer, indicating regions
of charge depletion and excess with respect to the superposition
of isolated layers. (c) Schematic double-layer geometry in top
view, with the cutting plane used in (a) and (b) indicated by the
dash-dotted line. \label{Fig1}}
\end{figure}

\begin{figure}[b]
\includegraphics[width=0.90\columnwidth]{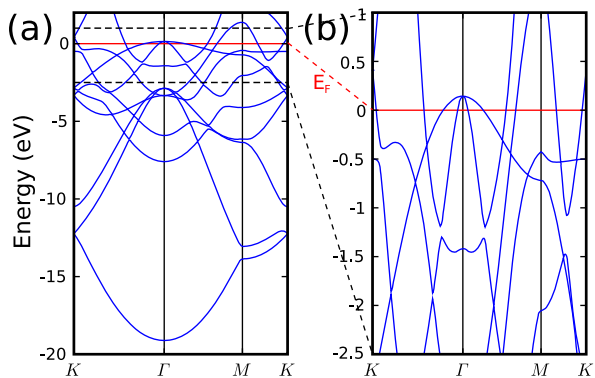}
\caption{(Color online) (a) Electronic band structure $E({k})$ of
a Pd monolayer interacting with a graphene layer. (b) Details of
$E({k})$ in (a) near the Fermi level, defined as $E_{\text{F}}=0$.
\label{Fig2}}
\end{figure}

To gain insight into the electronic structure in the contact
region, we performed density functional theory (DFT) calculations
of Ti and Pd monolayers interacting with a graphene layer. We
described valence electrons by Troullier-Martins pseudopotentials
and used the Perdew-Zunger form of the exchange-correlation
functional in the local density approximation (LDA) to DFT, as
implemented in the \textsc{SIESTA} code
\cite{soler-tsmfaioms2002}. With a double-zeta
basis and a 100~Ry energy cutoff in the plane-wave expansions of
the electron density and potential, we found the total energy to
be converged to ${\alt}1$~meV/atom. We performed a full structure
optimization to determine the equilibrium adsorption geometry, the
adsorption energy, and the local charge redistribution caused by
the metal-graphene interaction. Since the interatomic distances in
bulk Pd (2.7~{\AA}) and Ti (2.95~{\AA}) lie close to the honeycomb
spacing in graphene (2.46~{\AA}), we only considered epitaxial
adsorption. For both Pd and Ti, we found a slight preference for
the sixfold hollow site on graphite, depicted schematically in
Fig.~\ref{Fig1}(c). For Pd, we found the equilibrium inter-layer
distance to be 3.2~{\AA}, consistent with a relatively weak,
mostly covalent bond energy of 0.3~eV per Pd atom. The interaction
between an epitaxial Ti monolayer and graphene was only
insignificantly stronger with 0.4~eV per Ti atom at an inter-layer
distance of 3.0~{\AA}.

The quality of nanotube-electrode contacts has been shown to
depend sensitively on the Schottky-barrier in semiconducting
nanotubes~\cite{heinze-cnasbt2002} and band bending in metallic
nanotubes, both reflecting the charge transfer within the
junction. Our Mulliken population analysis indicates a net charge
transfer of only ${\approx}0.1$~electrons from Pd and Ti to the
graphene layer. More useful information is contained in the charge
redistribution, depicted in Fig.~\ref{Fig1}. Results for Pd
electrodes, shown in Fig.~\ref{Fig1}(a), suggest an accumulation
of excess charge in the region between Pd and graphene layers. As
seen in Fig.~\ref{Fig1}(b), the charge redistribution in
Ti/graphene is very different, suggesting charge accumulation in
the atomic layers and depopulation of the interlayer region and
thus an increase of the interlayer scattering potential. The lower
scattering potential and the populated interlayer state at the
Pd/C junction appear well suited for carrier injection into the
nanotube, making the Pd/C contact superior to the Ti/C contact.

Besides the charge transfer, the contact quality depends even more
sensitively on the nanotube-metal
hybridization~{\cite{knoch-aemfcnft2004}}, which is addressed in
Fig.~\ref{Fig2} for the Pd/graphene system. The density of states
at $E_{\rm F}$ assumes a large value, which is a
pre-requisite for a good contact. The electronic band structure of
the system, depicted in Fig.~\ref{Fig2}(a), suggests that all
states are closely related to either Pd or graphene states. In the
Pd/graphene system, the graphene bands are rigidly shifted by
${\Delta}E_{\text{C}}=0.374$~eV and the metal bands by
${\Delta}E_{\text{Pd}}=-0.020$~eV with respect to the isolated
layers. In the Ti/graphene system, the rigid band shift at the
junction is much stronger, ${\Delta}E_{\text{C}}=-1.15$~eV, and
has the opposite sign to Pd.

Especially interesting for the transparency of the contact is the
nature of Pd-C rehybridization, which is best visible in
Fig.~\ref{Fig2}(b) close to the Fermi level. Particularly clear is
the hybridization between Pd$d_{z^2}$ and C$p_z$ orbitals, which
causes a ${\approx}0.15$~eV band splitting about $0.5$~eV below
$E_{\rm F}$, in the vicinity of the $K$ point. Since $K$ denotes
also the Fermi momentum of graphene, this occurrence of Pd-C
hybridization near this $k$-point suggests an efficient way to
inject carriers into graphene near the Fermi energy without
involving phonons to conserve momentum.

To obtain quantitative information about the effect of the
junction geometry and hybridization on the transparency of the
contact, we performed large scale quantum transport calculations
of nanotubes in contact with metal electrodes within the Landauer
approach~{\cite{cuniberti-trocime2002}}. Our calculations for
nanotube segments exceeding $10^2$~nm were facilitated using an
efficient $O(N)$ decimation
algorithm~{\cite{PdCNT05-nemec-ext-cont}}. Our transport
calculations were based on a simplified tight-binding Hamiltonian,
describing only the interaction between Pd$d_{z^2}$ and C$p_z$
orbitals. We found that the electronic band structure of the
Pd/graphene system near $E_{\rm F}$, depicted in
Fig.~\ref{Fig2}(b), can be reproduced by considering the
hybridization between the $pp\pi$ band of graphene, associated
with $\gamma_0=2.66$~eV, and a Pd-based band, using
$t_{\text{Pd/C}} \approx 0.15$~eV for the hopping integral between
Pd and each of the six C neighbors. Such a simple mapping turned
out insufficient to describe the hybridization between Ti and
graphene. Based on typical band repulsion observed in that system,
the Ti-C hopping integral $t_{\text{Ti/C}} \approx 0.3$~eV should
be about twice that found for Pd.

\begin{figure}[t]
\includegraphics[width=0.80\columnwidth]{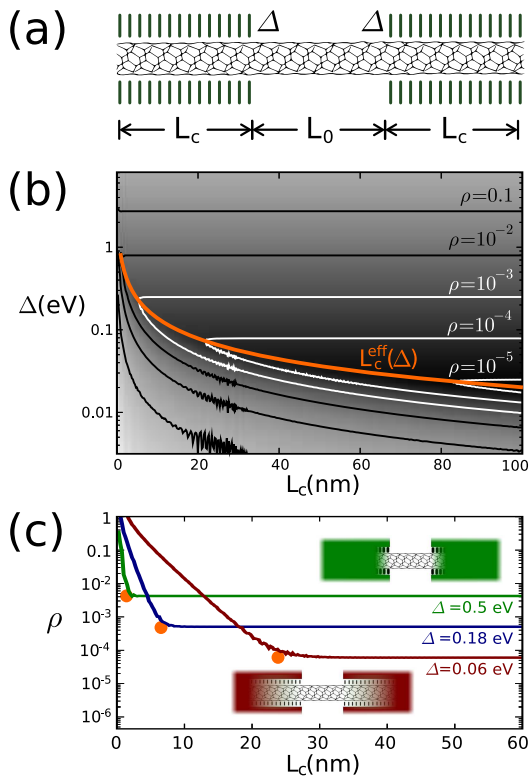}
\caption{(Color online) (a) Schematic geometry of the $(6,6)$
nanotube in contact with metal leads, used in the calculation of
the contact reflection coefficient $\rho$. (b) Contact reflection
coefficient $\rho$ as a function of the nanotube-metal coupling
$\Delta$ and the contact length $L_{\text{c}}$. (c) Cuts through
the contour plot (b) at selected values of $\Delta$, showing
$\rho$ as a function of $L_{\text{c}}$. The effective contact
length $L_{\text{c}}^{\text{eff}}$ is emphasized by a heavy solid
line in (b) and by data points in (c).\label{Fig3}}
\end{figure}

Results of our transport calculations for metal-nanotube junctions
are summarized in Fig.~\ref{Fig3}. In our schematic view of an
extended contact, depicted in Fig.~\ref{Fig3}(a), we distinguish
three regions within a finite tube. The central region of length
$L_0$, describing an unperturbed nanotube, is connected at both
ends to contact regions of varying length $L_{\rm c}$. In both
extended contact regions, each atom is coupled to the coating
metal electrode in a similar way, as previously considered in
Ref.~\cite{nakanishi-cbcname2000}.

In the model examined in the following, we chose {diagonal
wide-band leads}, which contacted each atom of the nanotube
independently. In this case, the coupling is described by an
energy-independent, purely imaginary self energy
$\Sigma(E)=-\text{i}\Delta$, where $\Delta=t_{\text{Me/C}}^2
\mathcal{N}_{\text{Me}}(E_{\rm{F}})$ can be extracted from our
{\em ab initio} results. Using the calculated density of states
$\mathcal{N}_{\text{Me}}(E_{\rm{F}})$ at Pd and Ti surfaces, which
is of the order $1$~eV$^{-1}$, we obtain
$\Delta_{\text{Pd}} \approx 0.06$~eV and
$\Delta_{\text{Ti}} \approx 0.3$~eV, assuming that each carbon
atom is in direct contact with three metal atoms.

Transmission $T$ through a molecular conductor is generally
limited by the number of channels $N_{\text{ch}}$, which depends
on the band structure of a perfectly periodic system as
$T \leq T_{\text{bands}}=N_{\text{ch}}$. The reason for the
effective transmission through a conductor with contacts
$T(L_{\text{c}},\Delta)$ being lower than through the ideal
infinite system $T_{\text{bands}}$ lies in the reflection at the
contacts. To quantify the quality of a nanotube contact, we define
the {contact reflection coefficient} by
\begin{equation}
  \rho(L_{\text{c}},\Delta)=
  \frac{1}{\left\langle{T(L_{\text{c}},\Delta)}\right\rangle}-
  \frac{1}{T_{\text{band}}} \label{defrho}\;.
\end{equation}
The average is taken around the Fermi level, in a region between
the first van-Hove singularities, similar to a transmission
convolution capturing hot electron
effects~{\cite{PdCNT05-alternative-rho}}. Our results in
Fig.~\ref{Fig3}, based on ensemble averaging in
Eq.~(\ref{defrho}), agree quantitatively with those obtained by
averaging the transmission using the Fermi distribution, for a
wide range of temperatures. Physically, the total resistance $R$
of such an idealized system could be separated as
$R=R_{\text{band}}+R_{\text{c}}$, where
$\left.{R_{\text{band}}}=1/(2G_0\right)$ is given by the quantum
limit of two open channels with conductance $G_0=2e^2/h$ each.
Assuming zero temperature and neglecting disorder effects, $R_{\rm
c} \sim \rho/G_0$ originates only from sub-optimal contacts and
may theoretically be arbitrarily small.

We generally expect zero transmission in the limiting cases of
vanishing contact length, $L_{\text{c}}=0$, and vanishing
coupling, $\Delta=0$. For finite values of $L_{\text{c}}$ and
$\Delta$, however, it is not obvious if a combination of strong
coupling and short contact is superior to a combination of weak
coupling and a long contact. To obtain a quantitative answer, we
calculated $\rho$ for a contact to a $(6,6)$ armchair nanotube
with $L_0=100$~nm as a function of $L_{\text{c}}$ and $\Delta$. We
found $\rho$ to be independent of the tube diameter, as long as
the energy range used in the averaging avoids subband-related van
Hove singularities. We also found no dependence on $L_0$, as long
as $L_0$ was much larger than a unit cell size. Consequently,
$\rho$ should be only a function of $L_{\rm c}$ and $\Delta$.


Our results for $\rho(L_{\text{c}},\Delta)$ are depicted in
Figs.~3(b) and (c). We find the contour plot of $\rho$ in
Fig.~\ref{Fig3}(b) separated into two regions by a line of
``minimum contact reflection''
$L_{\text{c}}^{\text{eff}}(\Delta)$. In short contacts, the
transparency is restricted by $L_{\text{c}}$ and $\rho$ increases
with decreasing $L_{\text{c}}$ due to a generalized Breit-Wigner
broadening of the resonances. The pronounced ripples found at
small values of $\Delta$ and $L_{\text{c}}$ are not numerical
artefacts, but rather reflect the interplay between resonances in
a finite nanotube segment and conduction electrons propagating
with the Fermi momentum. For very weak coupling, we find
$\left\langle{T}\right\rangle_E \propto L_{\rm c}\Delta$. For
larger values of $L_{\rm c}\Delta$, however, dissipation along the
contact region modifies this simple behavior, yielding
$\rho \propto \exp\left(-{\alpha}L_{\text{c}}\Delta\right)$.

In sufficiently long contacts, defined by
$L_{\text{c}}>L_{\text{c}}^{\text{eff}}$, $\rho$ becomes
independent of $L_{\rm c}$ and is given by $\rho=\rho_0\Delta^2$.
We found that our results can be reproduced well by using
$\rho_0=0.016$~eV$^{-2}$.
The independence of $\rho$ from $L_{\text{c}}$ in long contacts is
seen clearly in the plots of $\rho(L_{\text{c}})$, depicted in
Fig.~\ref{Fig3}(c) for selected values of $\Delta$. Particularly
intriguing appears our result that reflection is minimized in case
of a weaker specific coupling, provided the contact is
sufficiently long.

This physical origin of this unexpected behavior is schematically
illustrated in the insets in Fig.~\ref{Fig3}(c). Saturation
transparency is reached for relatively short contacts in the case
of strong coupling. In this case, however, the abrupt change in
the electronic structure between the uncoated and the coated
nanotube segment causes extra reflection. This change is less
abrupt in the case of weaker coupling and a long effective contact
region $L^{\text{eff}}_{\text{c}}$, reducing the overall
reflection.

To estimate the effective contact length
$L_{\text{c}}^{\text{eff}}$, we make use of the above described
expressions for $\rho$ in the adjacent regions in the
$(L_{\text{c}},\Delta)$ plane. The line, where these two functions
intersect, corresponds to the line of minimum $\rho$, and is given
by the analytical expression
\begin{equation}
  L_{\text{c}}^{\text{eff}}\left({\Delta}\right)=
  \ell_{\text{uc}}\frac{\alpha_1}{\Delta}\ln\frac{\alpha_2}{\Delta} \;.
\label{Eq2}
\end{equation}
Here, $\ell_{\text{uc}}=2.46$~{\AA} is the unit cell length. The
parameters $\alpha_1=1.34$~eV and $\alpha_2=9.14$~eV were obtained
by fitting our numerical data.
In the specific case of Pd and Ti, we found
$L_{\text{c}}^{\text{eff}}\left({\Delta_{\text{Pd/C}}}\right)
 \approx 30$~nm and
$L_{\text{c}}^{\text{eff}}\left({\Delta_{\text{Ti/C}}}\right)
 \approx 4$nm.

Since realistic metal-nanotube contacts are rarely epitaxial, we
have considered various forms of disorder in the contact region
and determined their effect on transport. We modelled weak to
moderate disorder by randomly perturbing the metal-nanotube
coupling $\Delta$ with respect to $\Delta_{\rm Me/C}$. We did not
find noticeable change in $\rho$ even for perturbations of
$\Delta$ as large as its reference value. As another extreme case,
we modelled strongly diluted contacts by randomly suppressing the
interaction between individual metal and nanotube atoms in the
contact region, down to one percent of contacts with respect to
the epitaxial case. Finally, we considered various forms of
non-diagonal contributions to the self-energy, modelling the
cross-coupling between neighboring metal atoms. In all the
studies, which addressed deviations from epitaxy at the interface
and our description of the leads, we found the same behavior as
depicted in Fig.~\ref{Fig3} and analytically described by
Eq.~(\ref{Eq2}), with possibly modified numerical values of
$\rho_0$, $\alpha_1$ and $\alpha_2$ \cite{PdCNT05-nemec-ext-cont}.%

Our main conclusion, which proved to be robust with respect to
variations in the details, is that each contact can be
characterized by an effective contact length
$L_{\text{c}}^{\text{eff}}$, which depends only on the local
metal-nanotube coupling, not on the diameter of the tube. Assuming
that the effective contact length between the nanotube and the
electrode exceeds $L_{\text{c}}^{\text{eff}}$, which is likely the
case in most experimental setups, then a higher contact
transparency is expected when the metal-nanotube coupling is
weak~\cite{nakanishi-cbcname2000}. Especially in very long
contacts, the sensitive $\rho = \rho_0 \Delta^2 \propto
t_{\text{Me/C}}^4$ dependence of $\rho$ on the coupling strength
$t_{\text{Me/C}}$ may be taken as an important guideline,
suggesting to minimize coupling and maximize contact length to
achieve a high contact transparency. In the specific case of Pd
and Ti contacts, the weaker nanotube-metal coupling in the case of
Pd is a good explanation for the superiority of Pd-based nanotube
contacts.

In conclusion, we combined {\em ab initio} density functional
theory with transport calculations to distinguish microscopically
between `good' and `poor' metal contacts to nanotubes. Comparing
Pd and Ti as examples of different contact metals, we traced back
the observed superiority of Pd to the nature of the metal-nanotube
hybridization. Based on large scale Landauer transport
calculations, we suggest that the `optimum' metal-nanotube contact
generally combines a weak hybridization with a large contact
length of typically few hundred nanometers between the metal and
the nanotube.

Of immediate interest is, of course, the general validity of our
results. We plan additional studies addressing the effect of
non-epitaxial contacts and the nature of charge
carriers~{\cite{PdCNT05-nemec-ext-cont}}. Particularly interesting
in this respect should be studying spin injection from
ferromagnetic contacts and Andreev reflection at the contact to
superconducting electrodes.

We acknowledge fruitful discussions with M.~Grifoni, A.~Nitzan,
J.~Fabian and C.~Strunk. This work has been funded by the
Volkswagen Stiftung under grant No.~I/78~340. DT was partially
supported by the Vielberth Foundation, the DFG-GRK 638, NSF-NIRT
grants DMR-0103587, ECS-0506309, and NSF NSEC Grant No.~425826.


\begin{thebibliography}{20}
\bibitem[{PdC()}]{PdCNT05-alternative-rho}
\bibinfo{note}{An alternative definition $\rho=\left\langle 1/T \right\rangle -
  1/T_{\text{band}}$ would differ from Eq.~(\ref{defrho}) only in the
  high-$\rho$ regime, which is not of primary interest here.}
\end{thebibliography}

\end{document}